\documentclass[reprint,amsmath,amssymb,aps,prresearch]{revtex4-2}

\usepackage{graphicx}
\usepackage{epstopdf}
\usepackage{amsmath,bm}
\usepackage{threeparttable}
\usepackage{multirow}
\usepackage{booktabs}
\usepackage{color}
\usepackage{hyperref}
\hypersetup{colorlinks=true,linkcolor=blue,filecolor=blue,urlcolor=blue,citecolor=blue,
}

\usepackage{float}

\begin{document}
	
	\title{Sliding Ferroelectrics Induced Hybrid-Order Topological Phase Transitions}
	\author{Ning-Jing Yang$^{1,2}$}%
	\author{Jian-Min Zhang$^{1,2}$}%
	\email[Corresponding author:]{ jmzhang@fjnu.edu.cn}
	\author{Xiao-Ping Li$^{3,4}$}%
	\email[Corresponding author:]{ xpli@imu.edu.cn} 
	\author{Zeying Zhang$^{5}$}
	\author{Zhi-Ming Yu$^{6,7}$}
	\author{Zhigao Huang$^{1,2}$}%
	\author{Yugui Yao$^{6,7}$}

	\affiliation{1 Fujian Provincial Key Laboratory of Quantum Manipulation and New Energy Materials, College of Physics and Energy, Fujian Normal University, Fuzhou 350117, China}
	\affiliation{2 Fujian Provincial Collaborative Innovation Center for Advanced High-Field Superconducting Materials and Engineering, Fuzhou, 350117, China}
	\affiliation{3 Research Center for Quantum Physics and Technologies, School of Physical Science and Technology, Inner Mongolia University, Hohhot 010021, China}
	\affiliation{4 Key Laboratory of Semiconductor Photovoltaic Technology and Energy Materials at Universities of Inner Mongolia Autonomous Region, Inner Mongolia University, Hohhot 010021, China}
	\affiliation{5 College of Mathematics and Physics, Beijing University of Chemical Technology, Beijing 100029, China}
	\affiliation{6 Centre for Quantum Physics, Key Laboratory of Advanced Optoelectronic Quantum Architecture and Measurement, School of Physics, Beijing Institute of Technology, Beijing 100081, China}
	\affiliation{7 Beijing Key Lab of Nanophotonics and Ultrafine Optoelectronic Systems, School of Physics, Beijing Institute of Technology, Beijing 100081, China}

	\begin{abstract}
		
		We propose ferroelectric layer sliding as a new approach to realize and manipulate topological quantum states in two-dimensional (2D) bilayer magnetic van der Waals materials. We show that stacking monolayer ferromagnetic topological states into layer-spin-locked bilayer antiferromagnetic structures, and introducing sliding ferroelectricity leads to asynchronous topological evolution of different layers (spins) owing to existence of polarization potentials, thereby giving rise to rich layer-resolved topological phases.  As a specific example, by means of a lattice model, we show that a bilayer magnetic 2D second order topological insulator (SOTI)  reveals an unrecognized spin-hybrid-order topological insulator after undergoing ferroelectric sliding. Interestingly, in such phase, the spin-up (top layer) and spin-down (bottom layer) channels exhibit first-order and second-order topological properties, respectively. Moreover, other topological phases such as SOTI, quantum spin Hall insulator, quantum anomalous Hall insulator, and trivial insulator can also emerge through changes in the parameters of the system, and the relevant topological indices are also discussed.  In terms of materials, based on first principles calculations, we predict material ScI$_{2}$ can serve as an ideal platform to realize our proposal. Further, we predict that the anomalous Nernst effect of these several topological phases exhibits distinct differences, and therefore can be used as a signal for experimentally probing.
	\end{abstract}
	
	\maketitle

	\textit{Introduction}---
	Topological phases and phase transitions play a central role in condensed matter physics and quantum technologies. Topological materials have evolved from traditional topological insulators to topological semimetals and topological crystalline insulators \cite{ RevModPhys.82.3045, RevModPhys.83.1057,hasan2011three, PhysRevLett.98.106803, burkov2016topological, RevModPhys.93.025002, PhysRevLett.106.106802, dziawa2012topological}, and more recently to magnetic and higher-order topological materials \cite{tokura2019magnetic,liu2023magnetic, PhysRevLett.122.206401, PhysRevLett.109.266405, hoti1, hoti2, hoti3, hoti4, hoti5, soti_ynj, PhysRevLett.130.116204}. 
	Recently, hybrid-order topological states were identified in phonon systems \cite{zhang2020, PhysRevLett.126.156801, PhysRevApplied.21.044002} and have also been observed in three-dimensional (3D) arsenic (As) electronic systems \cite{hossain2024}, while such states with hybrid orders have yet to be found in two-dimensional (2D) electronic materials.
	These materials typically exhibit distinct topological signatures such as surface, boundary, or corner states \cite{PhysRevB.86.081303, hoti2, sciadv18, gooth2016local, nature2018graphene, liu2024, ezawa2019, li2022robust}.
 	Controlling topological phase transitions has traditionally relied on external perturbations such as electric fields, magnetic fields, mechanical strain, and optical fields \cite{PhysRevLett.109.055502, weng2015quantum, PhysRevLett.130.196401, PhysRevLett.125.236805, tajkov2022revealing, 2021Light}. These methods typically induce band inversions by modifying the energy levels or symmetries of the electronic structure, allowing for precise tuning of topological phases.
	
	In contrast to external fields, intrinsic mechanisms capable of driving topological phase transitions are relatively rare. 
	Ferroelectricity, an intrinsic material property, has been explored in the context of topological phase transitions \cite{bai2025controllable, PhysRevB.108.125302, duan2024nonvolatile}.
	Recently, the role of interlayer ferroelectricity in 2D van der Waals (vdW) materials, especially induced by bilayer stacking sliding, has attracted considerable interest \cite{PhysRevLett.125.247601, fei2018ferroelectric, PhysRevLett.130.176801, gui2024stoner, PhysRevMaterials.8.024005, PhysRevB.110.024115, yang2024ferroelectric, liu2024tailoring, wu2021sliding, feng2023layer}. 
	Notably, interlayer sliding can be controlled experimentally, for instance, via Kelvin probe force microscopy (KPFM) \cite{wu2021sliding, meng2022sliding,sui2024atomic,vizner2021interfacial}.
	The topology of bilayer ferroelectricity has been discussed, with a focus on the layer Hall effect \cite{gao2024ferroelectric, tian2024quantum}. 
	Moreover, unlike conventional external fields, sliding ferroelectricity shifts energy levels \cite{PhysRevLett.125.247601}, potentially enabling new topological phenomena.
	Most studies focus on the symmetry changes induced by ferroelectric phase transitions and further to modulate topological phase transitions, while the role of polarization potential in the topological phase transition process has not been sufficiently considered. It is conceivable that polarization potential energy may play a significant role in the topological phase transitions beyond monolayer material systems, especially in sliding ferroelectric systems, where it may give rise to richer and more novel topological physics.

	\begin{figure}[h!]
		\centering
		\includegraphics[width= 7.5cm]{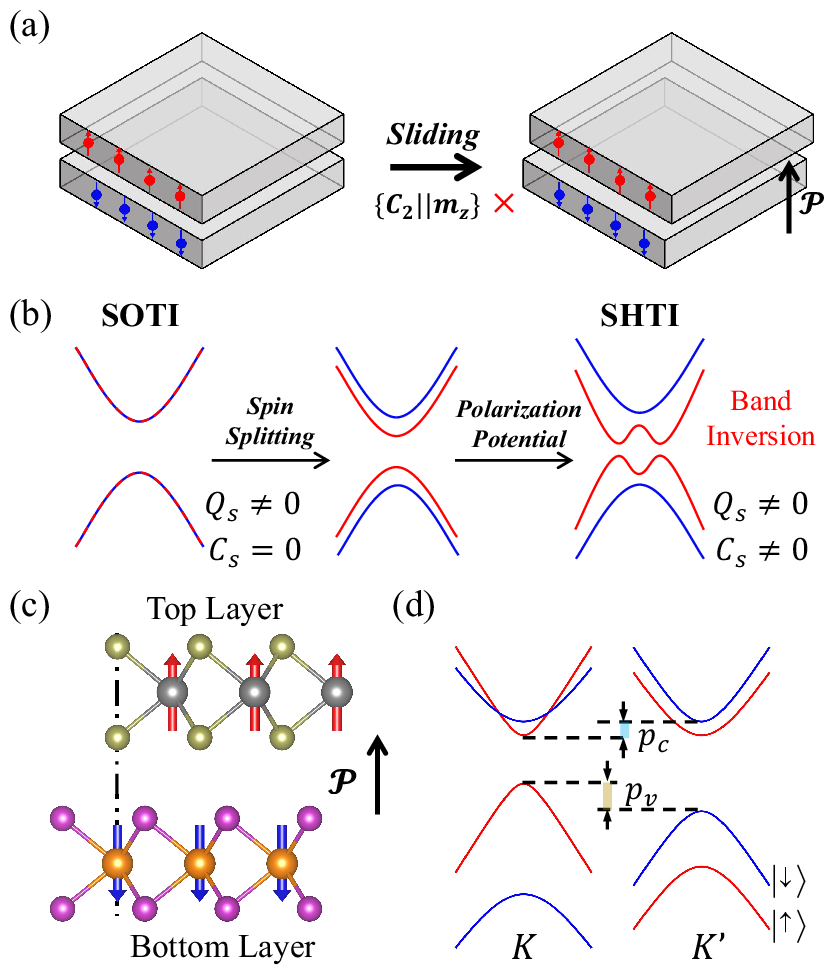}\caption{(a) The schematic diagram of bilayer AFM structures with sliding ferroelectricity. (b) The design of SHTI induced by interlayer sliding. (c) AB stacking configurations of the bilayer AFM lattice. (d) Band structure diagram for the bilayer AFM system under magnetoelectric coupling, illustrating valley ($K$ and $K'$) and spin degrees of freedom. The spin (valley) ferroelectric polarization splitting is marked with light brown and blue for the VBM ($p_v$) and CBM ($p_c$), respectively.}\label{pho:1}
	\end{figure}

	In this Letter, we introduce sliding ferroelectricity as a mechanism to induce topological phase transitions in antiferromagnetic (AFM) bilayer structures.
	Utilizing the spin- and layer-locking characteristics, we discuss the theoretical model of slide ferroelectricity and present the topological phase diagram. 
	We identify a novel spin hybrid-order topological insulator (SHTI), where the two spin components respectively exhibit first-order and second-order topology, arising from the asynchronous layer topological phase transition induced by polarization potential energy.  
	Additionally, the bilayer sliding system hosts both quantum spin Hall insulator (QSHI) and quantum anomalous Hall insulator (QAHI) phases during the phase transition.
	Through first-principles calculations, we demonstrate the SHTI results in the bilayer $\rm ScI_2$ system. Moreover, we also provide an experimental observation scheme utilizing the anomalous Nernst effect (ANE).
	Our research challenges the conventional understanding of topological phases by demonstrating that different topological orders can coexist within a single phase, enabled by sliding ferroelectricity. This highly tunable topological phase enables energy-efficient non-volatile memory and offers a pathway for robust quantum state manipulation in quantum technologies.

	\textit{Approach $\&$ Model}---
	We consider a spin- and layer-locked bilayer AFM second order topological insulator (SOTI). The interlayer sliding simultaneously generates ferroelectric polarization and breaks the \( \{C_2 || m_z\} \) symmetry, which consequently induces spin splitting, as illustrated in Fig. \ref{pho:1} (a).
	Fig. \ref{pho:1} (b) shows that SHTI arises from spin-selective band inversion driven by polarization potential during sliding. 
	The process starts from a SOTI with nonzero corner charge ($Q_s$) but  vanishing spin Chern number ($C_s$).
	Upon increasing polarization potential, a band inversion occurs in one spin channel, rendering both $Q_s$ and $C_s$ nonzero.
	The sliding ferroelectric polarization is also limited, as can be found in the general theory proposed by Ji \textit{et al.} \cite{PhysRevLett.130.146801}. Among the many 2D higher-order topological materials, systems with hexagonal lattices have a distinct advantage.

\begin{figure}
	\centering
	\includegraphics[width=7.5cm]{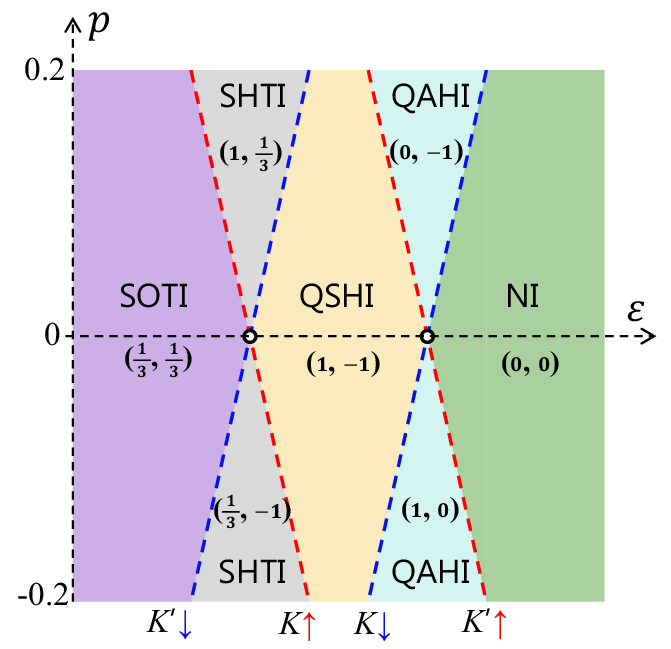}\caption{Topological phase diagram in the ($p$, $\varepsilon$) plane. The red and blue dashed lines represent phase boundaries indexed by the valley ($K, K'$) and spin ($S_z$) indices. The energy spectra of the four topological phases are shown in Fig. 3.}\label{pho:phase}
\end{figure}

\begin{figure*}
	\centering
	\includegraphics[width=15.5cm]{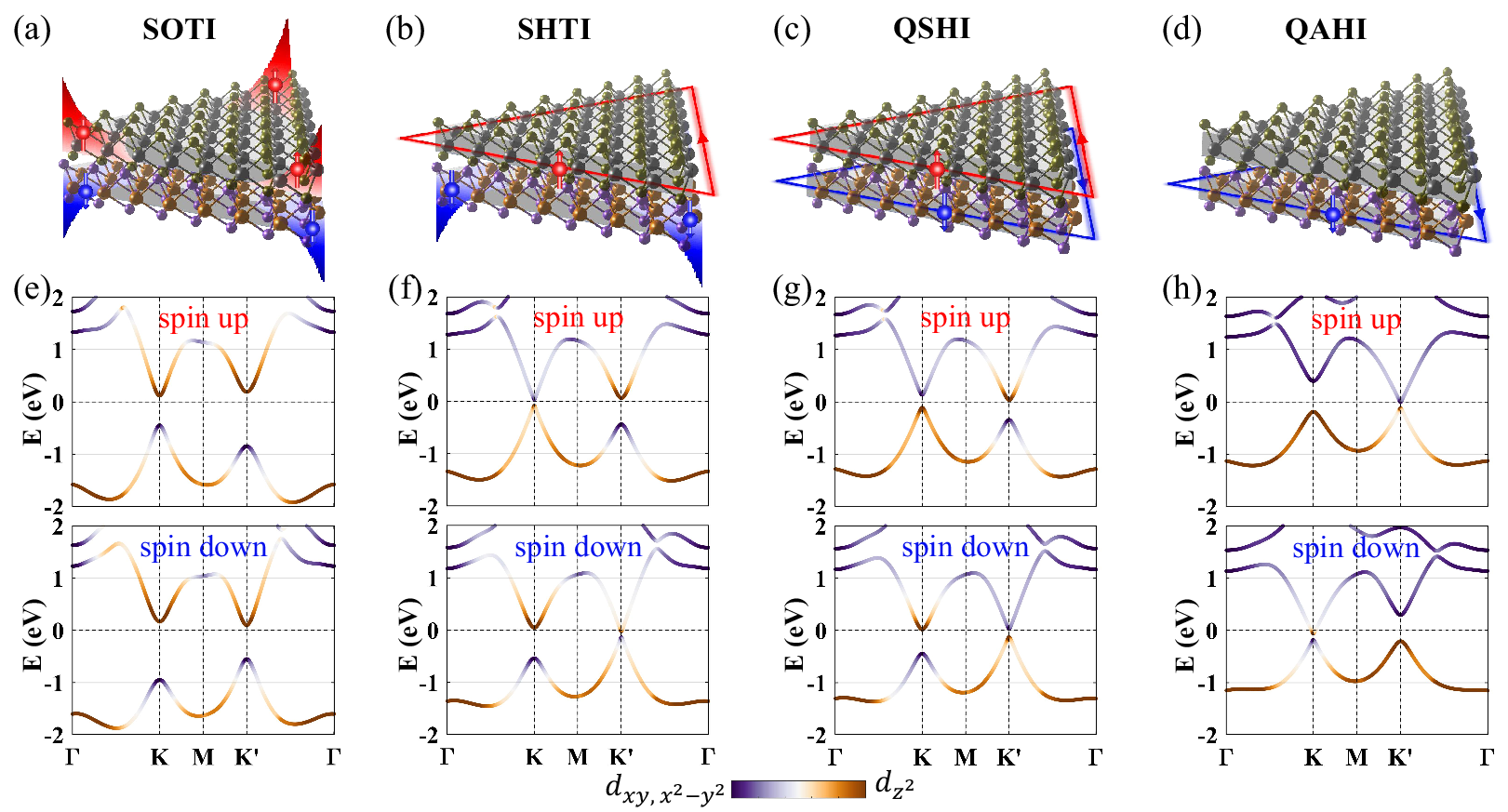}\caption{ 
		The real-space distribution schematics of the four spin-locked topological states, SOTI, SHTI, QSHI, and QAHI, are shown in (a)-(d), respectively. The corresponding spin-resolved three-band TB model band structures are shown in (e)-(h).}\label{pho:tb}
\end{figure*}

	We take the hexagonal bilayer MX$_2$ system shown in Fig. \ref{pho:1} (c) as a representative case for further analysis. Both the upper and lower layers consist of the same magnetic material, with different colors used to differentiate them. The magnetic atoms in the upper and lower layers exhibit spin-up and spin-down orientations, respectively. 
	In hexagonal systems, sliding ferroelectricity induces spin band splitting, particularly at the $K/K'$ valleys, as shown in Fig. \ref{pho:1} (d).

	We now use the tight-binding (TB) model to demonstrate the application of our method and explore how sliding induces SHTI. In AA-stacked bilayer AFM MX$_2$, the spin and layer degrees of freedom are intrinsically locked. Meanwhile, the AA-stacked bilayer AFM MX$_2$ system possesses $D_{3h}$ point-group symmetry, enabling its Hamiltonian to be described as a MoS$_2$-like three-band model based on the $d_{z^2}$, $d_{xy}$, and $d_{x^2-y^2}$ orbitals \cite{PhysRevB.88.085433}. By further incorporating the sliding ferroelectric term, the Hamiltonian can be expressed as:
	\begin{equation}
		\begin{split}	
			\mathcal{H} = &  \sum_{\langle ij\rangle}\sum_{\alpha\beta s}t_{ij}^{\alpha\beta}(1-\varepsilon) c_{i\alpha s}^{\dagger}c_{j\beta s}+\sum_{i\alpha s}\epsilon_{i\alpha}c_{i\alpha s}^{\dagger}c_{i\alpha s}
			\\ & +\sum_{i\alpha\beta s}\bm{\lambda_{\alpha\beta s}^{z}}c_{i\alpha s}^{\dagger}c_{i\beta s}+\sum_{i\alpha s}\bm{P_{\alpha s}}c_{i\alpha s}^{\dagger}c_{i\alpha s}+h.c. ,
		\end{split}
	\end{equation}
	where \(c_{i\alpha}^{\dagger}\) (\(c_{j\beta}\)) represents the electron creation (annihilation) operator for the orbital \(\alpha\) (\(\beta\)) at site \(i\) (\(j\)).
	\( \epsilon_{\alpha} \) denotes the on-site energy associated with each orbital, while \(t_{ij}^{\alpha \beta}\) represents the nearest-neighbor hopping energy between orbitals centered at sites \(i\) and \(j\). 
	$\varepsilon$ represents the magnitude of strain under tension, which governs the hopping strength.
	The third term describes the atomic intrinsic SOC, with \( \bm{\lambda_{\alpha\beta s}^z}=\lambda_I\bm{L_{\alpha\beta}^z}s_s^z \). The fourth term corresponds to the sliding ferroelectric contribution, \( \bm{ P_{\alpha s}} = \bm{P_{\alpha}}s_{z} \). 
	For further details on the TB model, refer to the supplementary materials \cite{sm}. 
	When sliding ferroelectric potential \( p=0 \), the system corresponds to AA stacking, where the bilayer AFM system belongs to the spin space group \( P^{-1}\text{-}6^{1}m^{-1}2^{\infty m}1 \). 
	In this system, upon sliding, the symmetry operation \( \{C_2 || m_z\} \) of the spin group is broken. In combination with the polar nature of the stacking configuration, this enables a ferroelectric polarization \cite{PhysRevLett.130.146801}.
	
	Since the spin interactions in this system are weak and the ferroelectric polarization potential decouples the spin components, spin Chern number can be used to characterize the topological properties of different spin channels, respectively. Thus, we first need to define a spin-hybrid-order topological index
	\begin{equation}
		\begin{split}	
			& I=(I^{\uparrow},I^{\downarrow}), I^{\uparrow/\downarrow}=Q^{\uparrow/\downarrow} \oplus C^{\uparrow/\downarrow},
		\end{split}
	\end{equation}
	where $Q^{\uparrow/\downarrow} $ and $C^{\uparrow/\downarrow}$ represent the spin corner charge and the spin Chern number, respectively. The system's second-order topological index $Q_s$ is protected by the $C_3$ rotational symmetry \cite{PhysRevB.99.245151}. For the parity calculation of all occupied states at high-symmetry points in the Brillouin zone, one can take $[\mathrm{K}_n^{(3)}]=\#\mathrm{K}_n^{(3)}-\#\Gamma_n^{(3)}$, where $\#$ denotes counting with respect to the symmetry eigenvalues at $K$ and $\Gamma$ points. The eigenvalues of the $C_3$ rotation are defined as $e^{2\pi i(n-1)/3}\left(\text{for} \ n=1,2,3\right)$. The topological index of higher-order topological crystalline insulators is $\chi^{(3)}=([K_1^{(3)}], [K_2^{(3)}]), \ Q_s^{(3)}=\frac{e}{3}[K_2^{(3)}]\text{mod}\ e$, where $e$ represents the charge of a free electron \cite{PhysRevB.99.245151, PhysRevLett.130.116204}. For spin-distinguished magnetic materials, spin corner charge $Q_s^{(3)} = e/3$.

	We construct a phase diagram in the $(p, \varepsilon)$ plane based on the topological index $I$ to characterize spin-resolved topological phases, as shown in Fig. \ref{pho:phase}. In the TB model, bilayer sliding is simulated via a ferroelectric polarization potential $p$. In the AA-stacked AFM bilayer (with $p=0$), tuning $\varepsilon$ drives concurrent topological transitions of spin-up and spin-down channels across the SOTI–QSHI–NI phases. However, in the AB-stacked configuration at the maximum polarization $p$, the symmetry breaking induces spin splitting, which leads to an asynchronous SOTI–QAHI–NI phase transition between spin-up and spin-down channels. Overall, the system exhibits a continuous evolution across five distinct phases: SOTI, SHTI, QSHI, QAHI, and normal insulator (NI), with corresponding topological indices $I_{SOTI}=(1/3, 1/3)$, $I_{SHTI}=(1, 1/3)$, $I_{QSHI}=(1, -1)$, $I_{QAHI}=(0, -1)$, and $I_{NI}=(0, 0)$. At the phase boundary, the system is a valley-polarized semimetal. In the BA configuration, these spin-hybrid-order indices are reversed.

	To compare different topological phases, Fig. \ref{pho:tb} shows band projection diagrams for four phases. Initially, the valence band at the $K$ valley is mainly from $d_{xy}$ and $d_{x^2-y^2}$ orbitals, characteristic of SOTIs before the topological phase transition \cite{soti_ynj}. In the triangular quantum dot under open boundary conditions, in-gap eigenvalues correspond to corner-localized states on both layers, as in Fig. \ref{pho:tb} (a). Band inversion first emerges at the $K$ valley for spin-up states as the nearest-neighbor hopping decreases, with the valence band maximum switching from $d_{xy}/d_{x^2 - y^2}$ to $d_{z^2}$ orbitals, as shown in Fig.~\ref{pho:tb}(f), rendering spin-up into a Chern topological insulator while spin-down remains a SOTI. The band-gap eigen wave functions reveal a SHTI, as shown in Fig. \ref{pho:tb} (b). Subsequently, spin-down undergoes a band inversion at the $K'$ valley. With Chern numbers of $+1$ and $-1$ for the upper and lower layers, respectively, the system enters the QSHI, as shown in Fig. \ref{pho:tb} (c). Another band inversion then occurs in the upper layer with $C_s=0$, driving the system into the QAHI, as shown in Fig. \ref{pho:tb} (d). Finally, as spin-down transitions into a NI, the system reaches a trivial phase.

	\begin{figure}
		\centering
		\includegraphics[width=7.5cm]{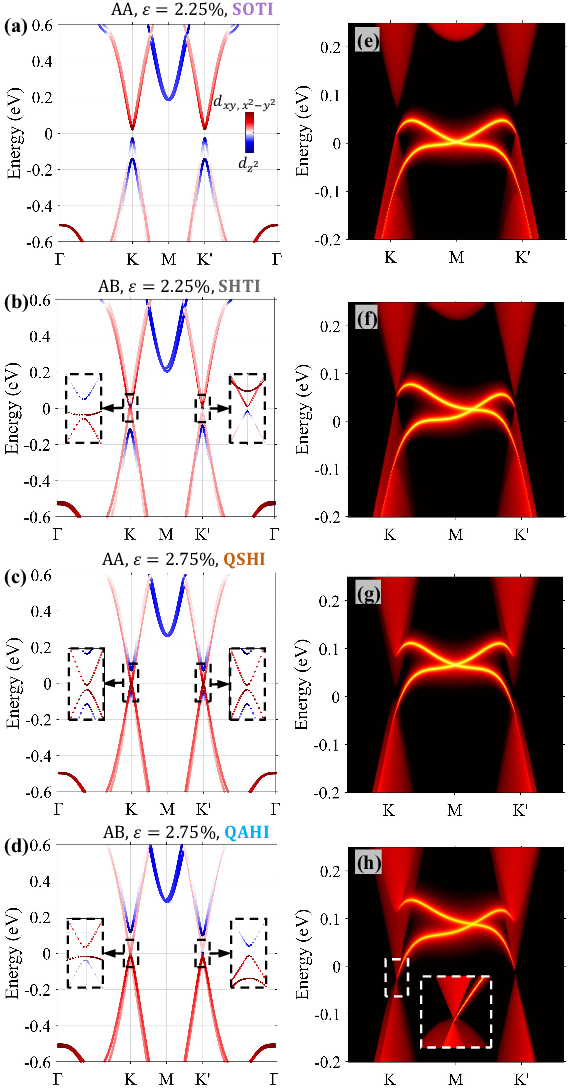}\caption{Under the modulation of tensile strain and sliding ferroelectricity, panels (a-d) depict the orbital-projected band structure of bilayer slide $\rm ScI_2$, while panels (e-h) show the surface states of the four topological phases.}\label{pho:dft2}
	\end{figure}

	\textit{Material realization}---
	The hybrid-order topological phase transition mechanism is general and applies to most 2D AFM higher-order topological bilayers. Sliding ferroelectric materials are often found in transition metal halides, such as MX$_2$ and MA$_2$Z$_4$ \cite{gao2024ferroelectric, tian2024quantum, li2023magnetic}. Compared to other tunable 2D magnetic materials like RuCl$_2$, FeCl$_2$, and VGe$_2$N$_4$, ScI$_2$ exhibits stronger interlayer ferroelectric polarization, with a valence band splitting of $p^v = 30 \ meV$ \cite{sm}.

    In the relaxed state, bilayer ScI$_2$ exhibits a corner charge $Q_c = 2e/3$, confirming it as a SOTI, with 2 representing the spin degree of freedom. At 2.25$\%$ strain in AA stacking, the band gap narrows, and the SOTI phase persists, as shown in Fig. \ref{pho:dft2} (a) and (e). Upon sliding to AB stacking, a first band inversion occurs at the $K$ valley, as shown in Fig. \ref{pho:dft2} (b). The sliding-induced symmetry breaking between opposite spin sublattices turns the bilayer into a compensated ferrimagnet rather than an AFM (see Fig. S10 \cite{sm}). Consequently, the surface band at $K$ valley hosts a single edge state connected to the conduction band, as shown in Fig. \ref{pho:dft2} (f). Increasing strain induces a second inversion at $K'$, yielding a QSHI at 2.75$\%$ strain with spin-polarized edge states, as shown in Fig. \ref{pho:dft2} (c, g). After sliding to the AB stacking, a third band inversion occurs at the $K'$ valley, driving the system into a QAHI with $C_s = -1$, as illustrated in Fig. \ref{pho:dft2} (d, h). Ultimately, further increasing strain drives the system into a NI via a final band inversion.

	\begin{figure}
		\centering
		\includegraphics[width=7.80cm]{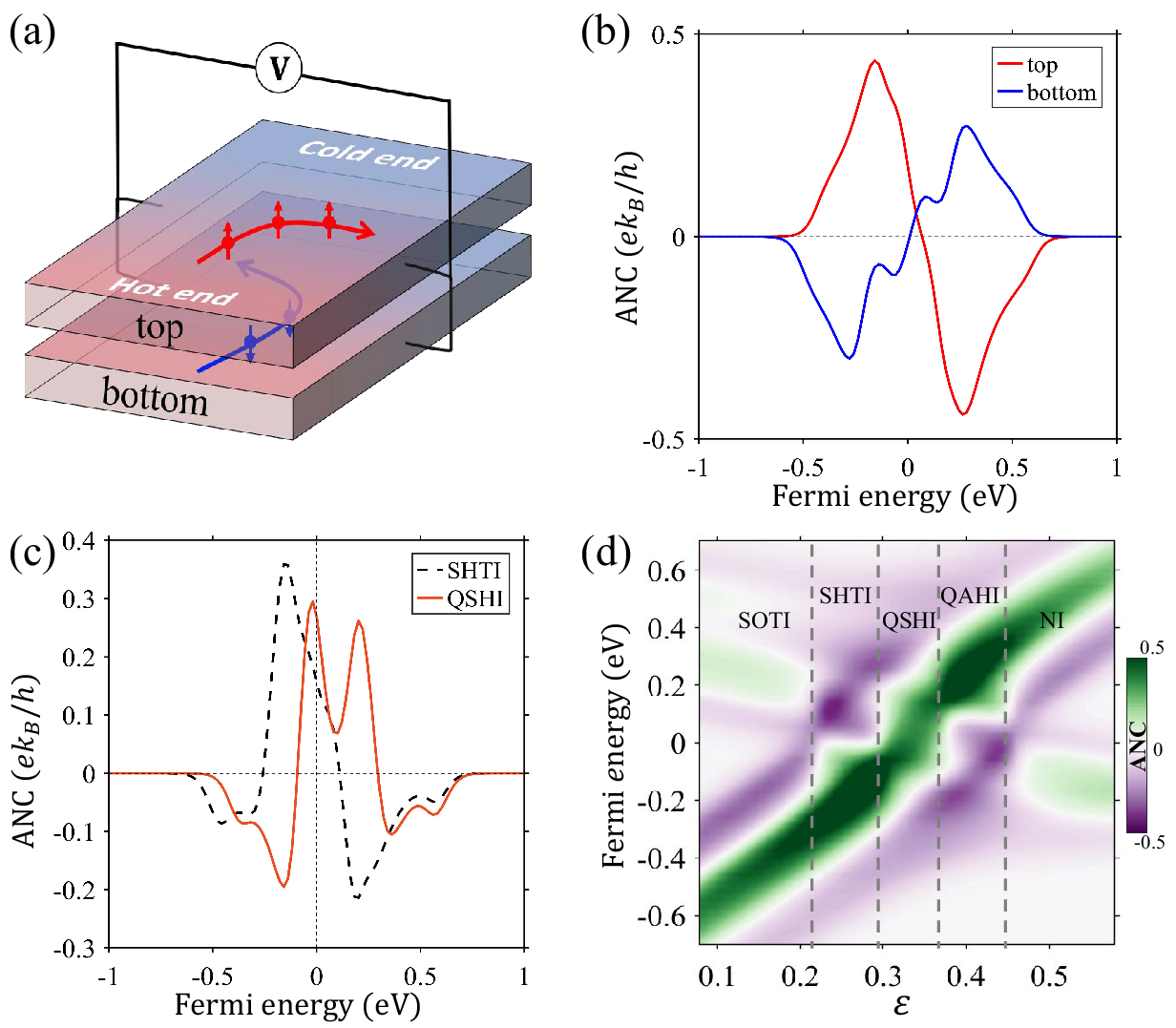}\caption{ 
			(a) Schematic of the Nernst conductivity in a bilayer sliding system under a temperature gradient $\nabla T$. (b) Spin-resolved ANC in the SHTI phase. (c) ANC of SHTI and QSHI during the sliding phase transition. (d) Topological phase diagram of ANC regulated by $\varepsilon$. We have set $\lambda = 0.2\ \rm eV$ and $p = 0.15\ \rm eV$.}\label{pho:ane}
	\end{figure}	

	\textit{Distinctive ANE}---
	The realization of SHTIs relies on higher-order topological phase transitions induced by sliding. In monolayer higher-order topological phase transitions, the ANE serves as a sensitive probe by directly reflecting the valley-resolved Berry curvature, which undergoes significant changes across phase transitions \cite{soti_ynj}. This sensitivity remains crucial in bilayer sliding ferroelectric systems, where the ANE effectively captures the evolution of topological phase transition. Moreover, the ANE offers an experimentally accessible thermal response, making it advantageous for device applications compared to conventional charge transport measurements. Therefore, we investigated the thermoelectric ANE characteristics at different stages to verify this phase transition process.  The anomalous Nernst conductivity is represented as:
	\begin{equation}
		\mathcal{N}=\frac{ek_{B}}{\hbar}\sum_n\int\frac{d^2k}{{(2\pi)}^2}\Omega^n(\mathbf{k})\mathcal{S}^n(\mathbf{k}),
	\end{equation}
	where $\mathcal{S}^n$ is the entropy density, $\Omega^n $ is Berry curvature. At 300 K temperature, we can obtain the anomalous charge Nernst conductivity (ANC): $ \rm{ANC} = \mathcal{N}_{top} + \mathcal{N}_{bottom}$. For more detailed discussions, please refer to the supplementary materials \cite{sm}.
	As illustrated in Fig. \ref{pho:ane}(a), which depicts the bilayer Nernst effect, the top and bottom layers are connected in parallel via probes, and the overall transverse electrical signal is subsequently measured. 
	After sliding, the Berry curvature in the top and bottom layers becomes inequivalent, leading to differences in the ANC output from the respective spin layers. Fig. \ref{pho:ane}(b) presents the spin-resolved ANC of the SHTI. The Berry-curvature-induced ANC changes across the topological phase transition due to band inversion. Fig. \ref{pho:ane}(c) further compares the ANC between the SHTI and QSHI phases.
	Near the Fermi level, the ANC evolves from a single peak to a double-peak, accompanied by a shift toward higher energies. Fig. \ref{pho:ane}(d) displays a staircase diagram of the ANC modulated by $\varepsilon$, where four clear topological phase boundaries, each corresponding to a band inversion, are observed. Thus, the five phases can be distinguished by their different ANC values. An advantage of the Nernst effect is that it is not constrained by low temperatures, making the observation conditions relatively accessible.

	\textit{Conclusions and Discussion}---
	In summary, we propose that sliding ferroelectricity enables asynchronous topological transitions between layers or spins, giving rise to layer-resolved topological phases, including the SHTI in 2D materials. This mechanism is confirmed in ScI$_2$, where we further suggest an ANE-based method to distinguish such phases. Ferroelectric polarization acts as a general and tunable tool for modulating layer-dependent topological states, with potential applicability to other layered materials. Our results indicate a route toward energy-efficient non-volatile memory and robust quantum state control. The emergence of hybrid-order topology hints at intrinsic mechanisms for complex topological phenomena, pointing to new directions in higher-order topological phase research.

	\vspace{3mm}
	This work is mainly supported by the National Natural Science Foundation of China (Nos. 11874113, 62474041, 12304086) and the Natural Science Foundation of Fujian Province of China (No. 2020J02018).

	\bibliography{References.bib}

\end{document}